\documentclass[preprint]{revtex4}
\usepackage{graphics}
\begin{document}
\title{Wealth distribution on complex networks}
\author{Takashi Ichinomiya}
\date{\today}
\affiliation{Department of Biomedical Informatics, Gifu University
Graduate School of Medicine, \\
Yanagido 1-1, Gifu 501-1194, Gifu, Japan}
\affiliation{PRESTO, Japan Science and Technology Agency}
\begin{abstract}
 We study the wealth distribution of the Bouchaud--M\'ezard (BM) model on complex
 networks. It has been known that this distribution depends on the topology of network by numerical simulations, however,  no one have succeeded to explain it.
  Using ``adiabatic'' and ``independent'' assumptions along
 with the central-limit theorem, we derive equations that determine the
 probability distribution function. The results are compared to those of simulations
 for various networks. We find good agreement between our theory and the
 simulations, except  the case of  Watts--Strogatz
 networks with a low rewiring rate, due to the breakdown of independent assumption.
\end{abstract}

\maketitle

\section{Introduction\label{Introduction}}

It is well known that the toplogy of the network changes the dynamics
dramatically. After the discovery of the absence of epidemic threshold
in a scale-free network\cite{Pastor-Satorras2001}, many researchers have
 focused on the dynamics on complex network, such as synchronization
\cite{Nishikawa2003, Ichinomiya2004}, pattern formation\cite{Nakao2008},
 and other phenomena.

In this paper, we study  the Bouchaud-M\'ezard (BM)
 model on complex network\cite{Bouchaud2000}.  
The power-law behavior of wealth distribution, Pareto's
distribution, has for over a century been one of the main
cornerstones of econophysics\cite{Pareto}. One of the simplest
models that explain this law is that proposed by Bouchaud and
M\'ezard, which consists of multiplicative noise
and globally coupled diffusion. After the proposal of the
BM model, several researchers numerically
investigated the generalized BM model in which diffusion occurs
between adjacent nodes in a complex network\cite{Garlaschelli2004,
Souma2001}. While, the research revealed that the network topology
alters the wealth distribution, there has been no quantitative
theory that sufficiently explains these simulation results.

Recently, we proposed a new theory for the BM model on a random
network\cite{Ichinomiya2012}. Using several assumptions that we
describe in the next section, we derived equations that determine
the static probability density function (PDF) of the wealth. The
results of this analysis were compared with those of the numerical
simulations and good agreement was obtained.

However, the wealth distributions were analyzed only on a random
network in the previous paper, and the distributions on other
complex networks was left as an open problem. The aim of this paper
is develop our previous work so that it is applicable to a general
complex network. Using the same techniques applied in the previous
paper, we derive the equations that determine the static PDF for the
BM model on a complex network for a given adjacency matrix. The
results are evaluated by a comparison with the numerical simulation
results, and our method is revealed to perform well for many network
systems.

This paper is organized as follows. In section \ref{Theory}, we
derive the self-consistent equations that determine the PDF of the
BM model on a complex network whose adjacency matrix is given. The
results obtained with this PDF are tested in sec. \ref{Simulation}
by comparing with simulations on several networks such as a random
network, the Watts--Strogatz (WS) network, and a real social
network. Finally, we discuss the results and future problems and
summarize the paper in sec. IV.

\section{Theory\label{Theory}}

We consider the BM model on a undirected complex network consisting
of $N$ nodes, whose adjacency matrix is given by $A=(a_{ij})$. The
dynamics of $x_i$, i.e., the wealth on node $i$, are determined by
the following Ito-type stochastic differential equations.
\begin{equation}
 dx_i = J\sum_{j=1}^N a_{ij} (x_j-x_i) dt + \sqrt{2}\sigma dW_i,\label{083834_8Aug12}
\end{equation}
where $J$, $\sigma$ and $W_i$ represents the diffusion constant of
wealth, strength of noise, and the standard Brownian motion, respectively.
This equation has no static PDF; however, the normalized  wealth
$x/\langle x \rangle$, where $\langle \cdots \rangle$ represents an
average over all nodes, can have a static PDF. For simplicity, $x$
is used in place of $x/\langle x  \rangle$ in the following
discussion.

To obtain a static $x_i$ distribution, we make the ``adiabatic and
independent'' approximation introduced in our previous
paper\cite{Ichinomiya2012}. We first assume that all $x_i$ values
are independent and that the correlation between different nodes is
negligible. Under this assumption, we assume the PDF %$\rho(x_1,
%x_2,\cdots , x_N)$
can be decomposed as $\rho(x_1,x_2,\cdots,
 x_N)=\rho_1(x_1)\rho_2(x_2)\cdots \rho_N(x_N)$.  Second, we assume that the rate of change of
$\bar x_i =\frac{1}{d_i}\sum_j a_{ij}x_j$, the average $x$ around
node $i$, is much slower than that of $x_i$, where $d_i$ represents
the degree of node $i$. Using these assumptions, we can calculate
the static PDF as follows.

Suppose $\bar{x}_i$ is constant. Then  $\rho_i(x|\bar{x}_i)$,  the
conditional PDF of $x$ on node $i$ for a given $\bar{x}_i$, is a
solution of the following Fokker--Planck equation.

\begin{equation}
 \frac{\partial \rho_i}{\partial t} = -\frac{\partial}{\partial
  x}[d_i(\bar{x}_i-x)\rho_i] +\sigma^2 \frac{\partial^2}{\partial x^2}
  [x^2 \rho_i]\;.
\end{equation}

The static solution of this equation, $\rho_{eq,i}(x|\bar{x}_i)$, is
given by
\begin{equation}
 \rho_{eq,i}(x|\bar x_i)=C_{d_i}(\bar x_i)\exp(-\alpha_{d_i} \bar x_i/x)x^{-2-\alpha_{d_i}}\;,\label{164143_7Aug12}
\end{equation}
where $\alpha_d =Jd/\sigma^2$ and  $C_d(x) = (\alpha_d
 x)^{1+\alpha_d}/\Gamma(1+\alpha_d)$. Using the adiabatic assumption, the static PDF of $x$ at node $i$ is given by
\begin{equation}
 \rho_{eq,i}(x) = \int d\bar{x} P_i(\bar{x})\rho_{eq,i}(x | \bar x)\;,\label{164204_7Aug12}
\end{equation}
where $P_i(\bar x)$ represents the PDF of $\bar x_i$.

To obtain $P_i(\bar x)$, we use the independent assumption again
together with the central-limit theorem. Under this assumption,
$P_i(\bar x)$ can be assumed to be the distribution of the average
of the independent variables $x_j$, where $j$ represents a node
adjacent to node $i$. If the variances of $x_j$ are finite for all
$j$, we can apply the central-limit theorem to approximate $P_i(\bar
x)$ by the Gaussian distribution
\begin{equation}
 P_i(\bar x ) = \frac{1}{\sqrt{2\pi} S_i} \exp\left[-\frac{(\bar x -M_i)^2}{2S_i^2}\right]\;,\label{091257_8Aug12}
\end{equation}
where $M_i$ and $S_i$ represent the average and variance of $\bar
x_i$, respectively. From Lindeberg's theorem, these values are given
by
\begin{equation}
 M_i = \sum_j a_{ij}\mu_j/d_i\;,\label{090008_8Aug12}
\end{equation}
and
\begin{equation}
S_i^2 = \sum_j a_{ij} s_j^2/d_i^2\;,\label{090021_8Aug12}
\end{equation}
where $\mu_j$ and $s_j^2$ are the average and variance of $x_j$,
respectively.

We can determine $M_i$ and $S_i$ by calculating $\mu_i$ and $s_i^2$
for $x$ using Eqs. (\ref{164143_7Aug12}), (\ref{164204_7Aug12}),
 (\ref{090008_8Aug12}), and (\ref{090021_8Aug12}). By using
 $\int_0^{\infty} dx \rho_{eq,i}(x|\tilde x) x = \tilde x$ and
$\int_0^{\infty} dx \rho_{eq,i}(x|\tilde x) x^2  =
\frac{\alpha_{d_i}}{\alpha_{d_i}-1}\tilde x^2$ for $\alpha > 1 $, we
find
\begin{equation}
 \int_0^\infty dx x\rho_i(x) = \int_{0}^{\infty} d\bar{x} P_i(\bar{x})
  \int_0^\infty dx x \rho_i(x|\bar{x}) =\int_{0}^{\infty} d\bar{x} \bar
  x P_i(\bar{x}) \sim M_i\;,\label{170529_22Aug12}
\end{equation}
and
\begin{equation}
  \int dx x^2 \rho_{i}(x) = \int_0^{\infty} d\bar x
  \frac{\alpha_{d_i}}{\alpha_{d_i}-1} \tilde x^2 P_i(\bar x) \sim
  \frac{\alpha_{d_i}}{\alpha_{d_i}-1} \left(M_i+S_i^2\right)\;,\label{170556_22Aug12}
\end{equation}
where we assume $M_i \gg \sqrt{S_i}$. We thus obtain $M_i=\mu_i=1$
for all $i$, and $S_i^2$ is the solution of
\begin{equation}
 s_i^2=\frac{\alpha_{d_i} S_i^2 + 1}{\alpha_{d_i} - 1}\;.\label{133408_24Aug12}
\end{equation}
Using Eq. (\ref{090021_8Aug12}), we can rewrite this equation as
\begin{equation}
 s_i^2=\frac{1}{\alpha_{d_i}-1} \left(1+\frac{\alpha_{d_i}}{d_i^2}\sum_j a_{ij} s_j^2\right)\;.\label{175036_7Aug12}
\end{equation}
For convenience and in anticipation for later analysis, we define a
new matrix $R=(r_{ij})$ as
\begin{equation}
 r_{ij}= \frac{\alpha_{d_i}}{d_i^2(\alpha_{d_i}-1)} a_{ij}\;.\label{defOfR}
\end{equation}
We then finally obtain the following equation for $s_i^2$:
\begin{equation}
 \sum_j (\delta_{ij}-r_{ij})s^2_j = 1/(\alpha_{d_i}-1)\label{eqForS}\;.
\end{equation}
Eqs. (\ref{164143_7Aug12}), (\ref{164204_7Aug12}),
(\ref{091257_8Aug12}), and (\ref{eqForS}) define the PDF of $x$.

Before concluding this section, we note in the following about wealth
condensation, defined as the divergence of $s_i^2$. Because $s_i^2
\ge 0$ for all $i$, Eq. (\ref{eqForS}) must have a solution
$s_j^2>0$ to have any meaningful result. If $J$ is large enough,
this condition is satisfied provided that $\mbox{min}(d_i)\ge 2$.
For a $J$ value that satisfies $\alpha_{d_i}>1$ for all $i$,
$R_{ij}$ is a non-negative matrix. Therefore, the solution of Eq.
(\ref{eqForS}) satisfies $s_i^2
>0$ if $\rho(R) <1$, where $\rho(R)$ represents the spectral radius
of $R$. From Frobenius's theorem, $\rho(R)$ is equal to the largest
eigenvalue of $R$, and thus we conclude that Eq. (\ref{eqForS}) has
a meaningful solution if the largest eigenvalue of $R$ is less than
1. Noting that $d_i = \sum_{j} a_{ij}$,  $\rho(R) \le
\mbox{max}(\sum_j |r_{ij}|)$, and $r_{ij} \rightarrow a_{ij}/d_i^2$
for $J\rightarrow \infty$, $\rho(R)$ is less than 1 for large $J$,
and a non-condensed phase appears, provided that $\mbox{min}(d_i)
\ge 2$.

In contrast, we always have a condensed phase at small $J$ due to
the divergence of the $r_{ij}$ elements. As we decrease $J$ from a
large value, the $r_{ij}$  values increase and diverge when
$\alpha_{d_i} = 1$; $\rho(R)$ also diverges in this case. Therefore,
the wealth is condensed if $J/\sigma^2  \le \mbox{max}(1/d_i)$.

% Finally we should note the important difference between our previous
% work and this one. In the previous work we studied the BM-model on
% random network with given degree distribution $Q(k)$. Therefore the
% result obtained is on the average over ensembles of random network,
% which have same degree distribution. On the other hand, what we derived
% in this paper is the wealth distribution on network with a adjacency
% matrix $a_{ij}$. Therefore our result must be tested using single
% network, not on the average over ensembles on networks.

\section{Simulations on various network models\label{Simulation}}

To test our theory, we compare the results with numerical simulation
results on several networks. We begin with the network created by
the Erd\"os-R\'enyi algorithm and then consider several networks
created by the WS model\cite{Watts1998} to test the effect of
clustering. It is shown that our theory fails when the rewiring rate
$p$ is too small, but good agreement is obtained when $p\ge 0.1$.
Finally, we test our theory on a real network. It would be ideal to
apply the theory to a real economic network, but as there is
currently no available data, we use the American college football
network obtained by Girvan and Newman\cite{Girvan} instead. For this
case, the PDF obtained from the simulations shows good agreement
with our theory. For all the simulations, $\sigma^2$ is set to 1.

\subsection{Random network}

\begin{figure}[t]
 \resizebox{!}{!}{\includegraphics{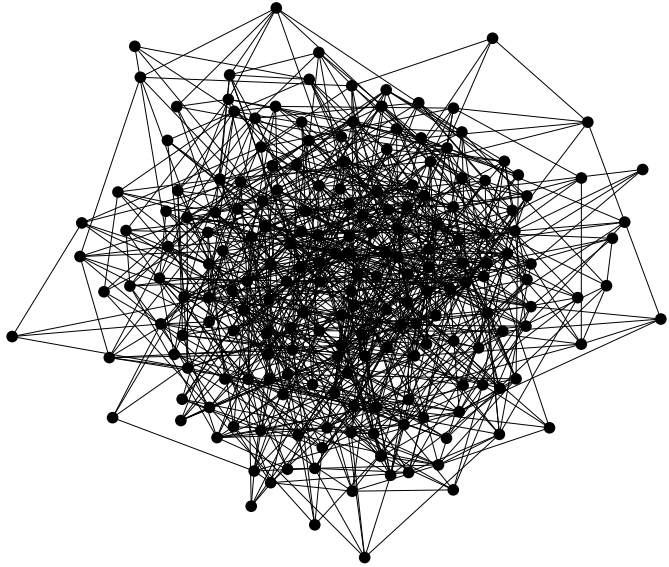}}
 \caption{The random network used for the simulations in sec. III.A}
 \label{140327_24Aug12}
\end{figure}

The network created by the Erd\"os-R\'enyi algorithm, shown in Fig.
\ref{140327_24Aug12}, has 200 nodes and a mean degree of $\langle k
\rangle = 10$.
\begin{figure}[t]
 \resizebox{.3\textwidth}{!}{\includegraphics{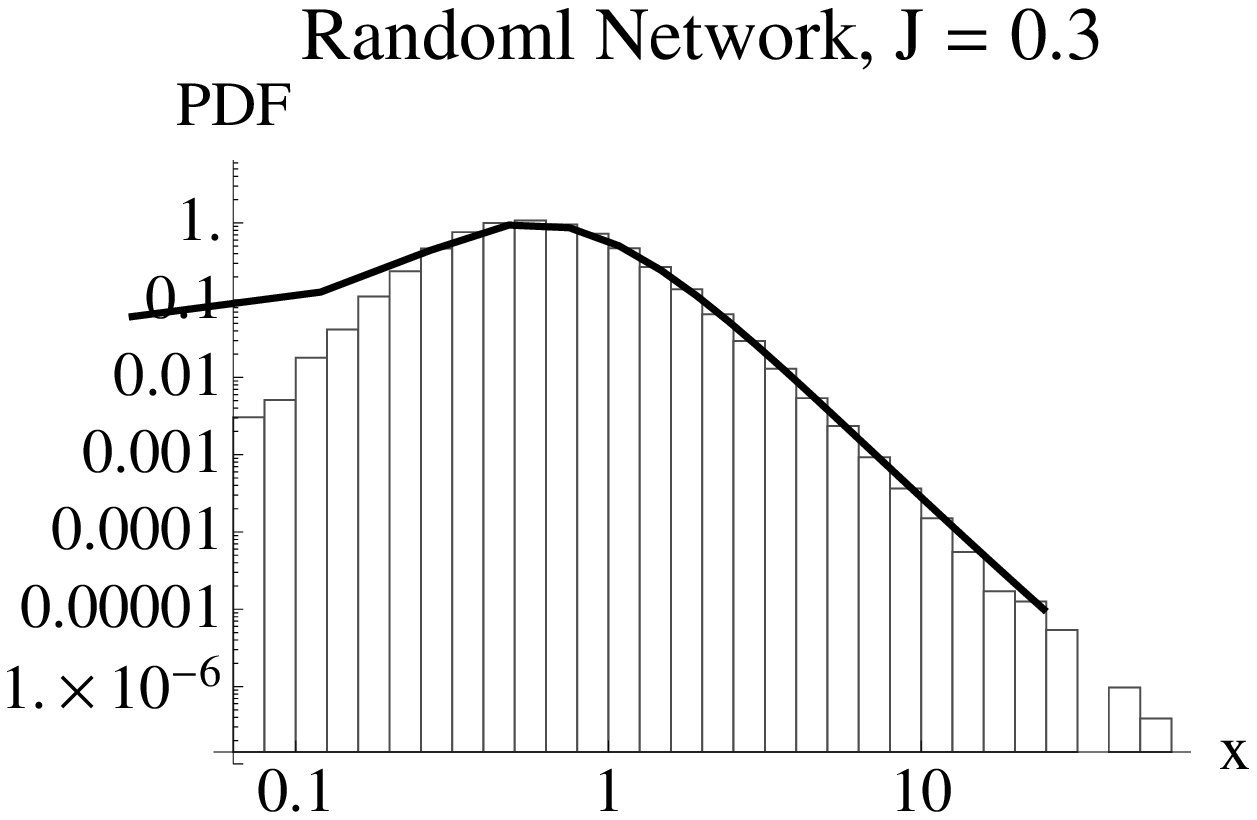}}
 \resizebox{.3\textwidth}{!}{\includegraphics{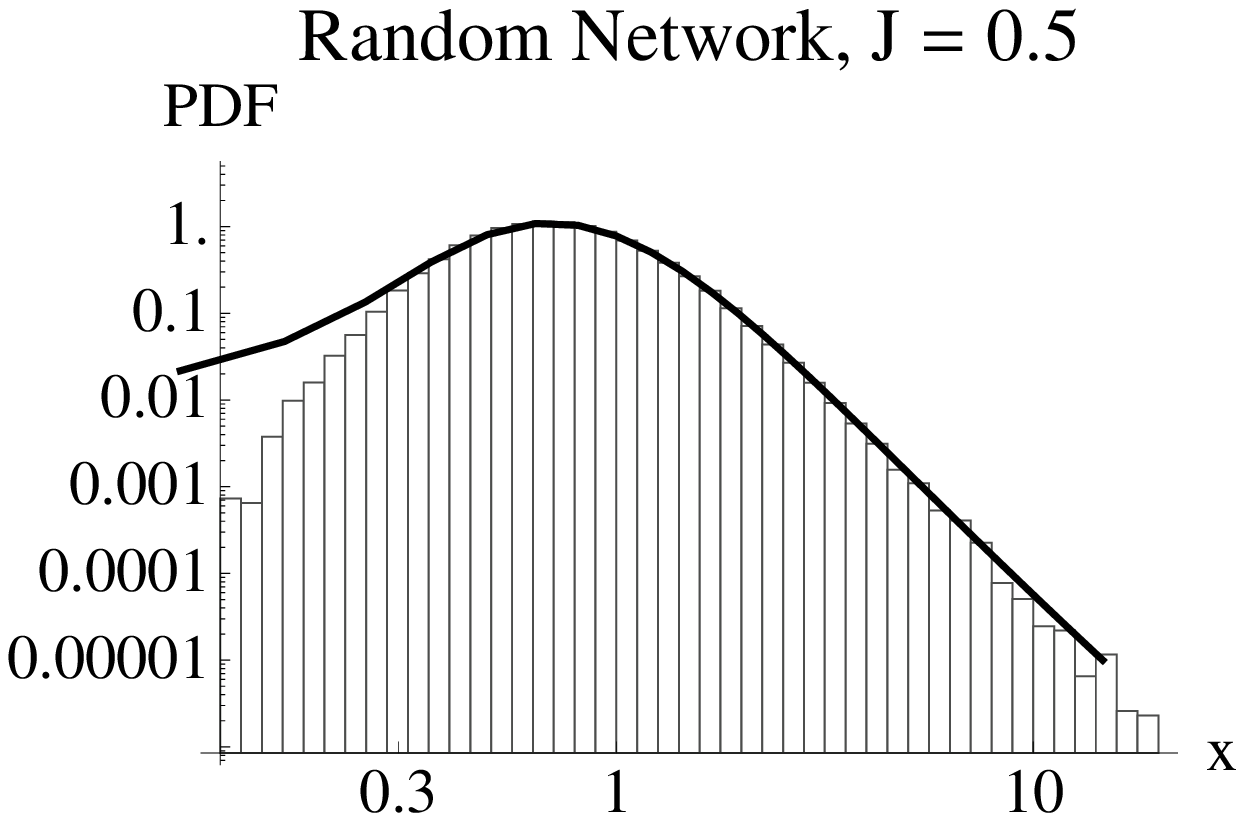}}
 \resizebox{.3\textwidth}{!}{\includegraphics{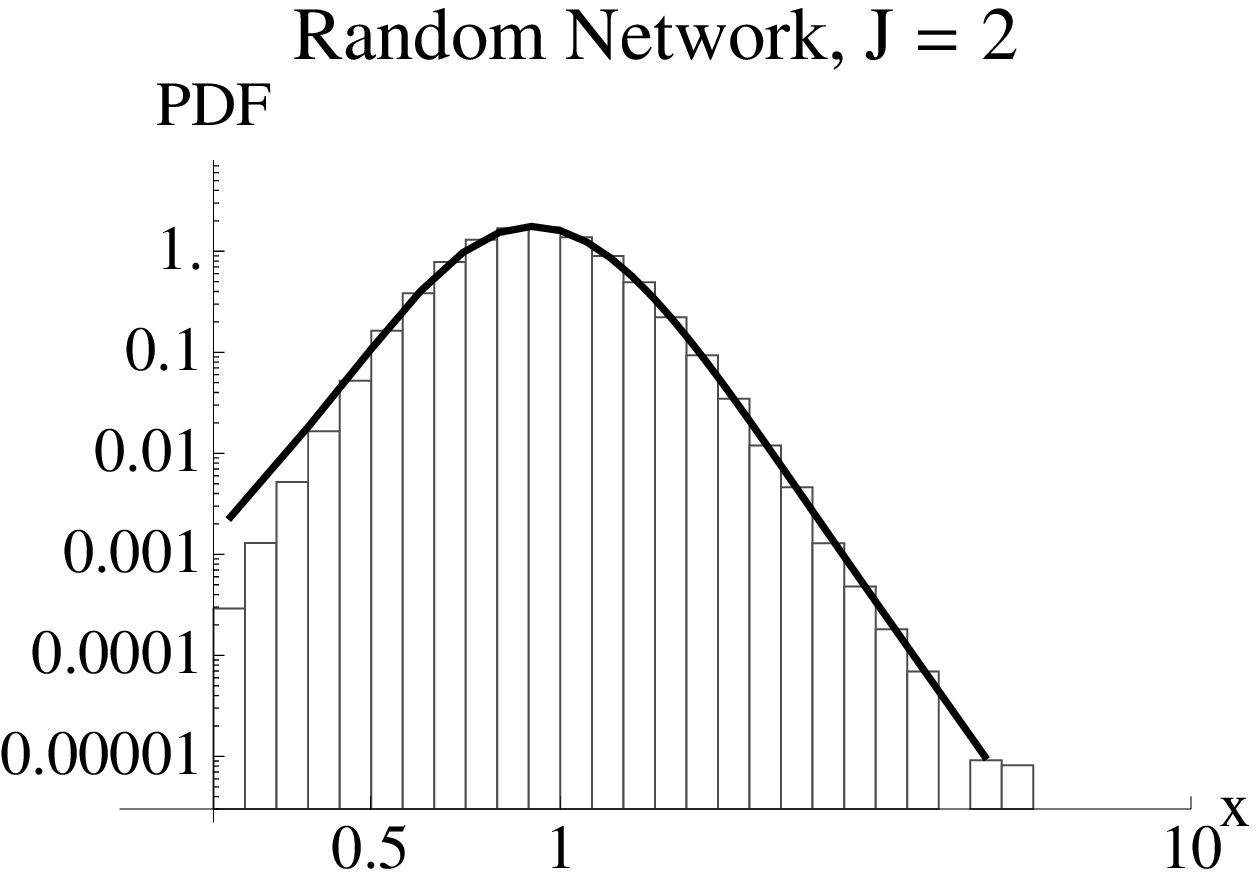}}
 \caption{A log-log plot of the PDF on the random network obtained by numerical simulation
 for $J=0.3$ (left), 0.5 (middle), and 2.0(right). }
 \label{170117_22Aug12}
\end{figure}

In Fig. \ref{170117_22Aug12} we plot the PDFs obtained by numerical
simulation for $J=0.3, 0.5$, and 2.0. The PDFs obtained using our
theory are indicated by the solid line and show good agreement with
the numerical simulation for all $x$ for the $J=2.0$ case, but there
are minor discrepancies at small $x$ values in the cases of $J=0.3$
and 0.5.
% Because the correlation
% of $x$ in the adjacent nodes will increase as $J$ increases, this
% discrepancy is not due to the breakdown of ``independent'' assumption.
The discrepancy is due to the $Mi \gg
 \sqrt{S_i}$ assumption used for deriving Eqs. (\ref{170529_22Aug12}) and (\ref{170556_22Aug12}).
The maximum $S_i^2$ value calculated by Eqs. (\ref{eqForS}) and
(\ref{133408_24Aug12}) is 0.33 for $J=0.3$ and 0.09 for $J=0.5$,
while $M_i = 1$. Thus, the $M_i \gg S_i$ approximation is not valid
as $\sqrt{S_i}$ is larger than 0.3. However, the approximation is
valid for the $J=2.0$ case as the maximum $S_i^2$ value is 0.01.

\subsection{Small-World Network}

\begin{figure}[t]
 \resizebox{.3\textwidth}{!}{\includegraphics{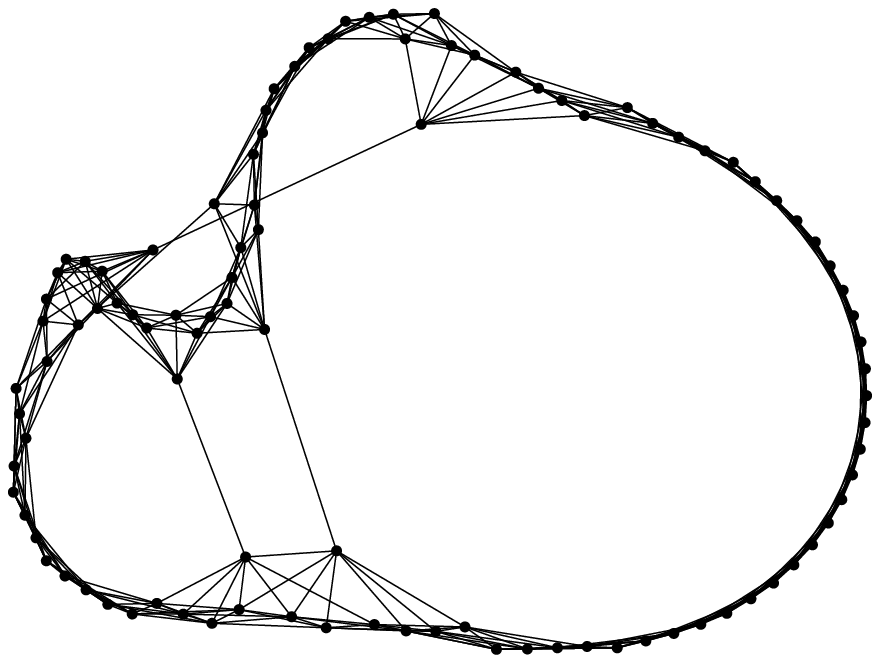}}
 \resizebox{.3\textwidth}{!}{\includegraphics{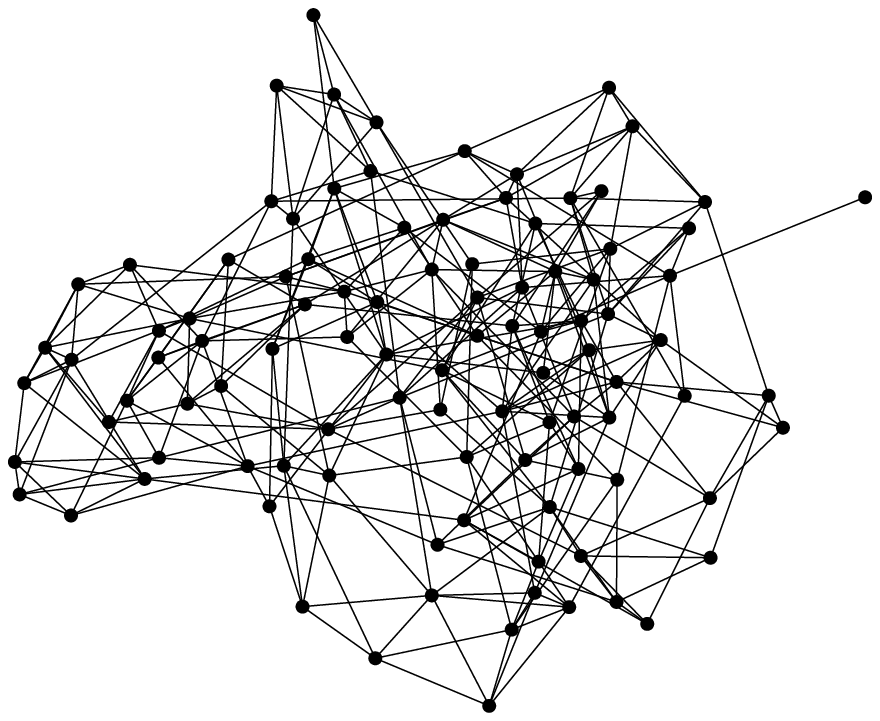}}
 \resizebox{.3\textwidth}{!}{\includegraphics{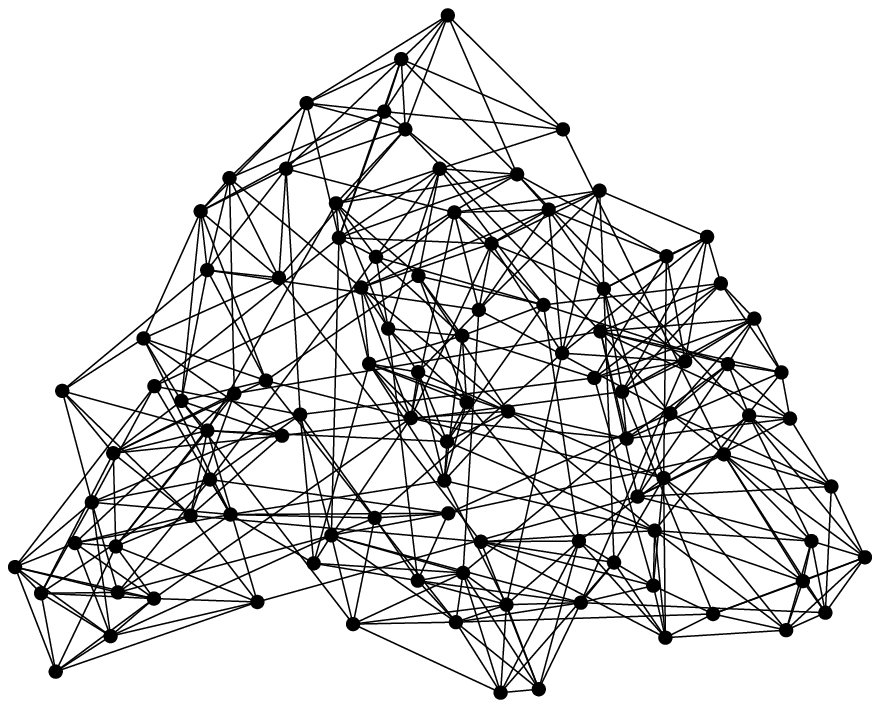}}\\
\resizebox{.3\textwidth}{!}{\includegraphics{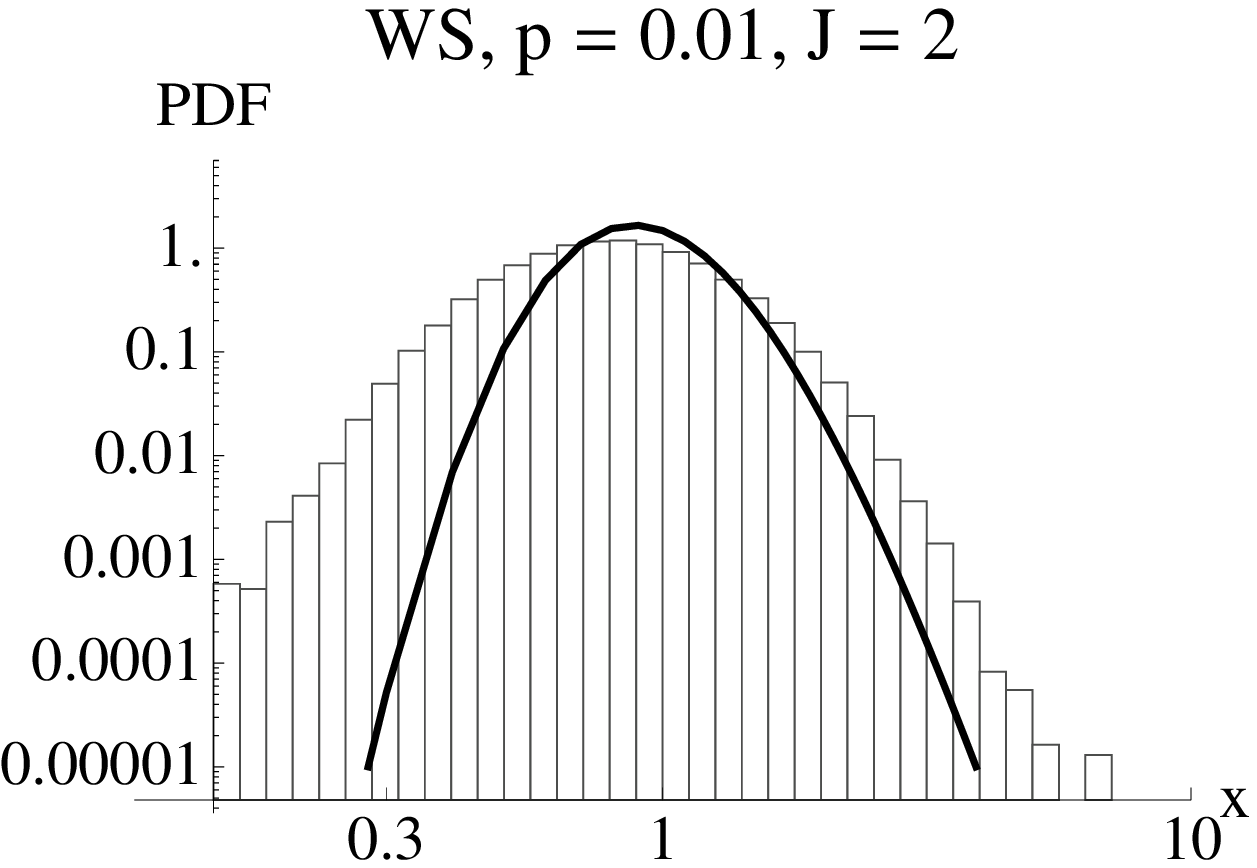}}
\resizebox{.3\textwidth}{!}{\includegraphics{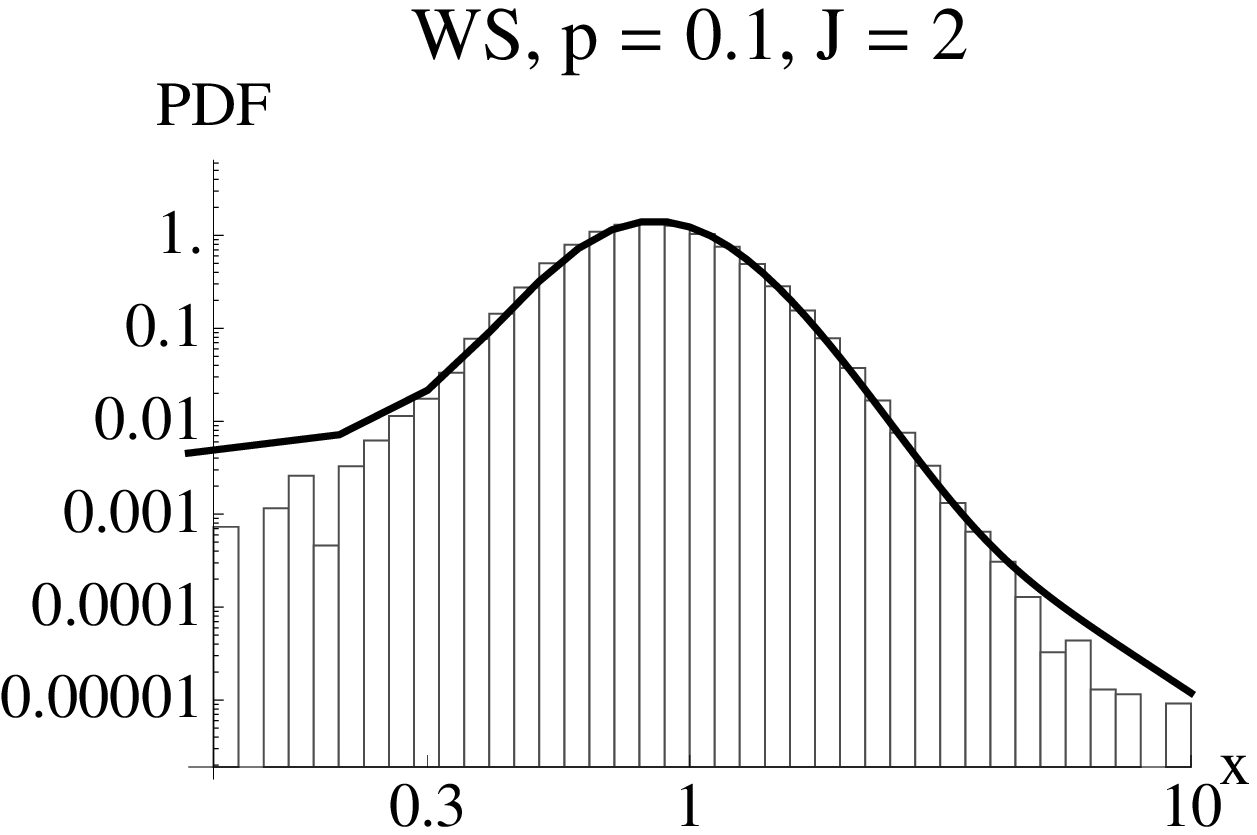}}
\resizebox{.3\textwidth}{!}{\includegraphics{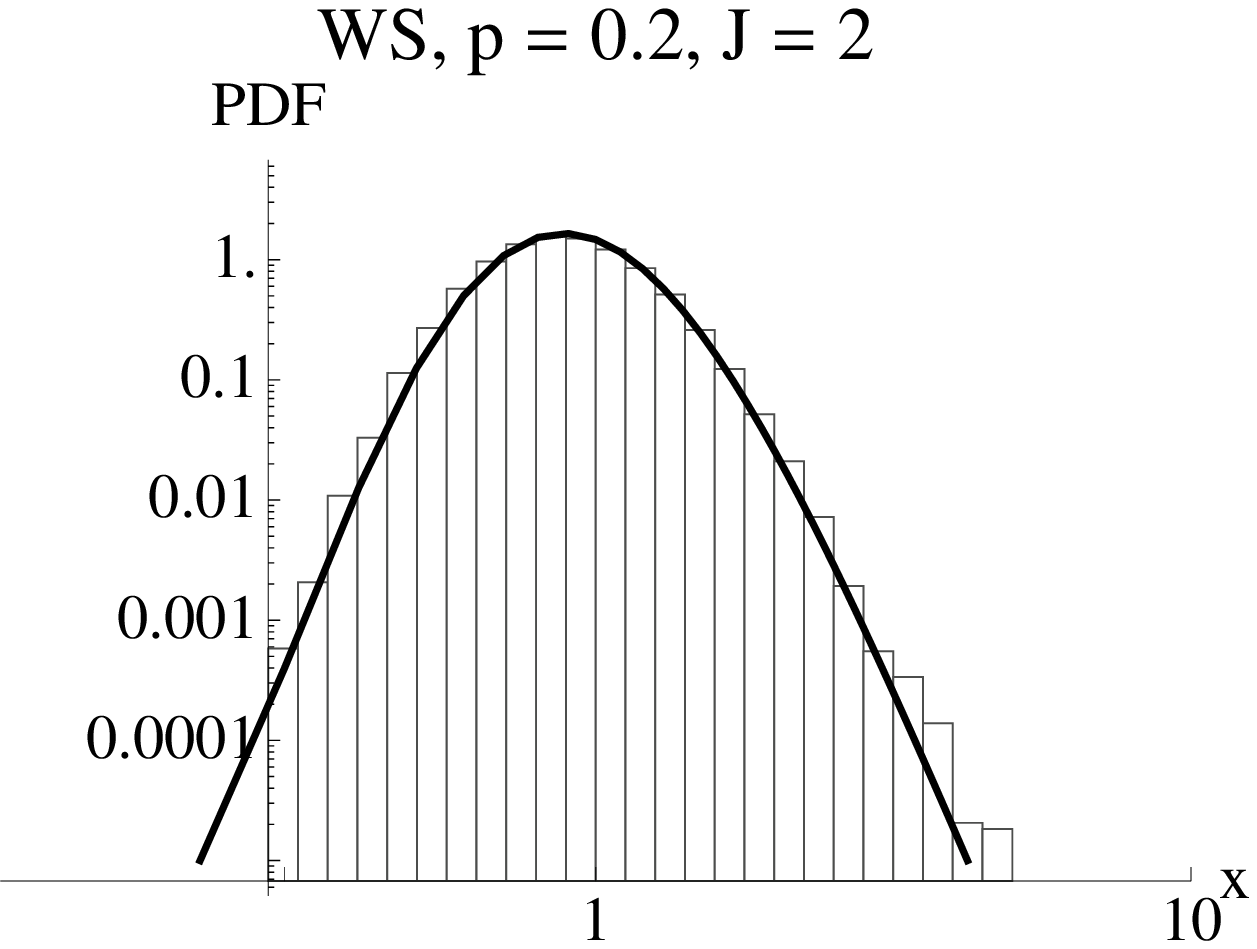}}
\caption{Upper: The small-world networks created by the
Watts--Strogatz algorithm with rewiring probabilities of
$p=0.01,0.1,$ and 0.2 used to test our theory. Lower: The PDFs
obtained from simulations on each networks with $J=2.0$. The solid
lines show the results of the method proposed here.}
\label{155309_15Aug12}
\end{figure}

Next, we consider three networks created by the WS algorithm. The
networks have $N=100$ nodes and a mean degree of $\langle d \rangle
= 8$ with different rewiring rates of $p=0.01, 0.1$, and 0.2. Fig.
\ref{155309_15Aug12} shows both the networks and PDFs. In the case
of $p=0.01$, our theory differs from the numerical simulation, while
we find good agreement for the $p=0.1$ and $p=0.2$ cases.

\begin{figure}[t]
 \resizebox{.45\textwidth}{!}{\includegraphics{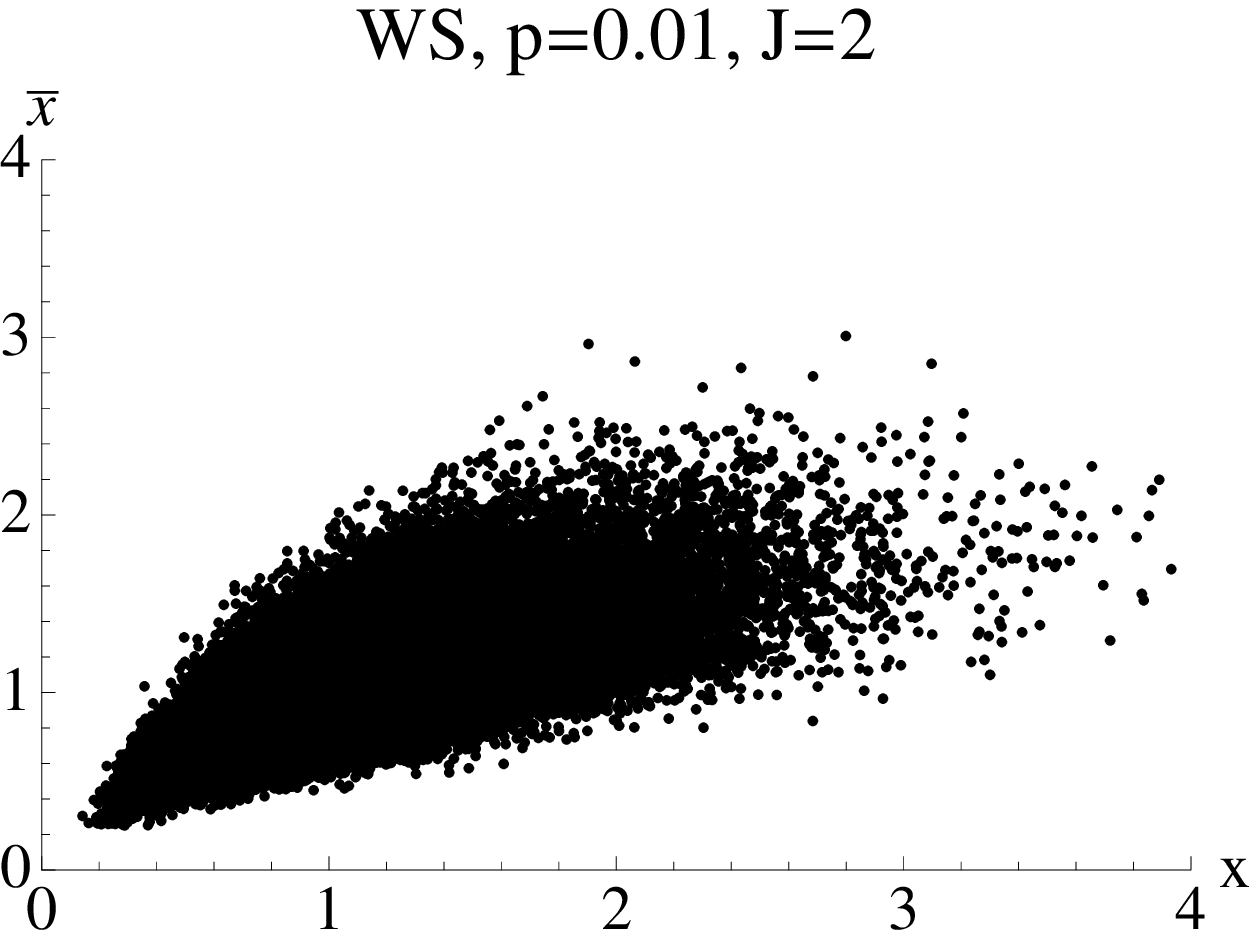}}
 \resizebox{.45\textwidth}{!}{\includegraphics{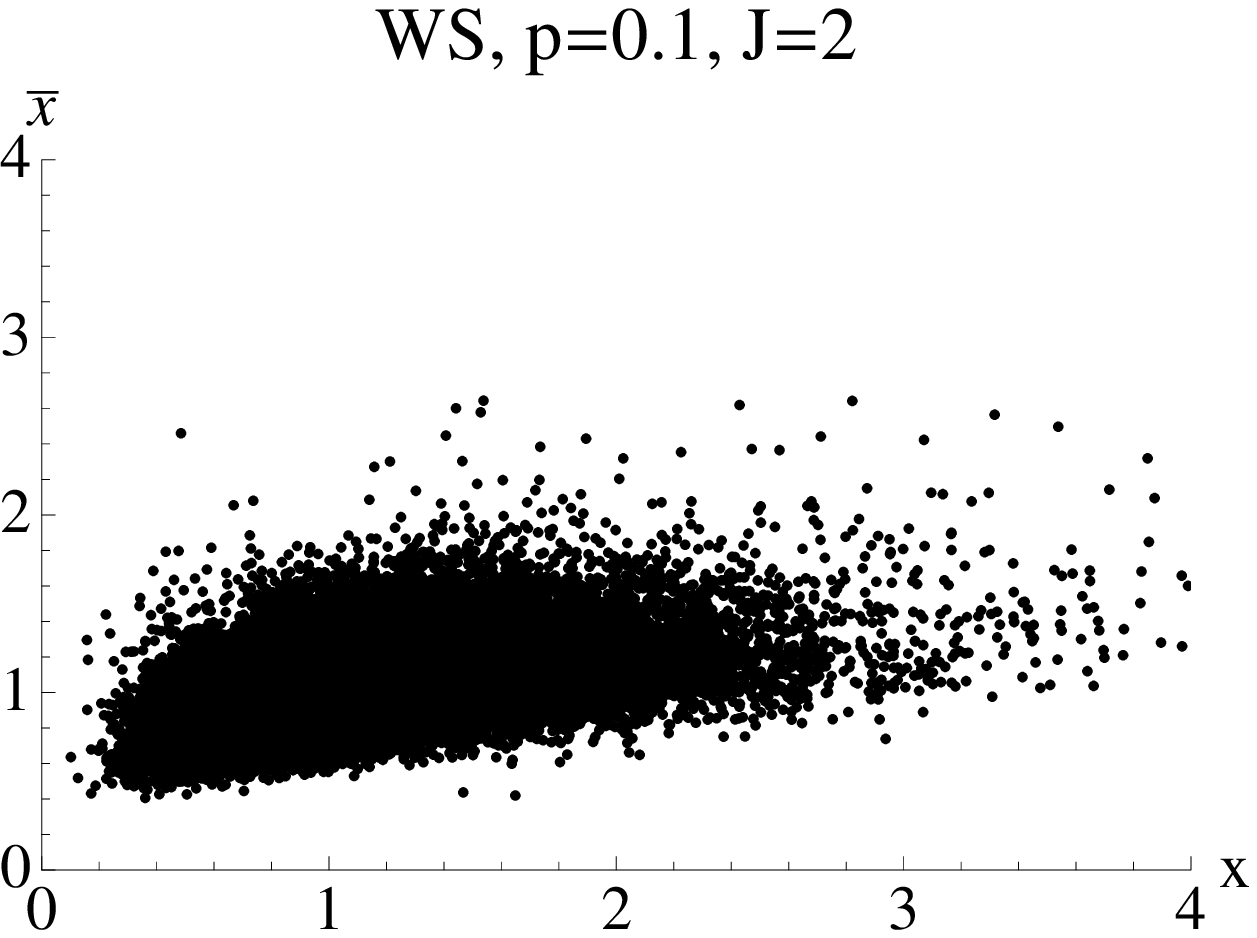}}
 \caption {Scatter plots of $x$ and $\bar x$ for the BM model on WS
 networks with $p=0.01$ and 0.1 and with $J=2$.}
 \label{scatterPlot}
\end{figure}

To investigate the reason for this discrepancy, we examine the
scatter plots of $x$ and $\bar x$ for $p=0.01$ and $p=0.1$ in Fig.
\ref{scatterPlot}. For $p=0.01$, we find that there is a strong
correlation between $x$ and $\bar x$ that weakens for $p=0.1$. The
differences can be quantified by the correlation, which is 0.74 for
$p=0.01$ and 0.48 for $p=0.1$. This result suggests that the discrepancy 
is caused by the breakdown of independent assumption. 

It is reasonable to question here when and why the independent
assumption fails. The small-world property does not appear to affect
the independence of wealth because the networks show high-clustering
and small betweenness for both the $p=0.01$ and 0.1 cases. We return
to this point in the discussion section.

\subsection{American college football network}

\begin{figure}[t]
 \resizebox{.45\textwidth}{!}{\includegraphics{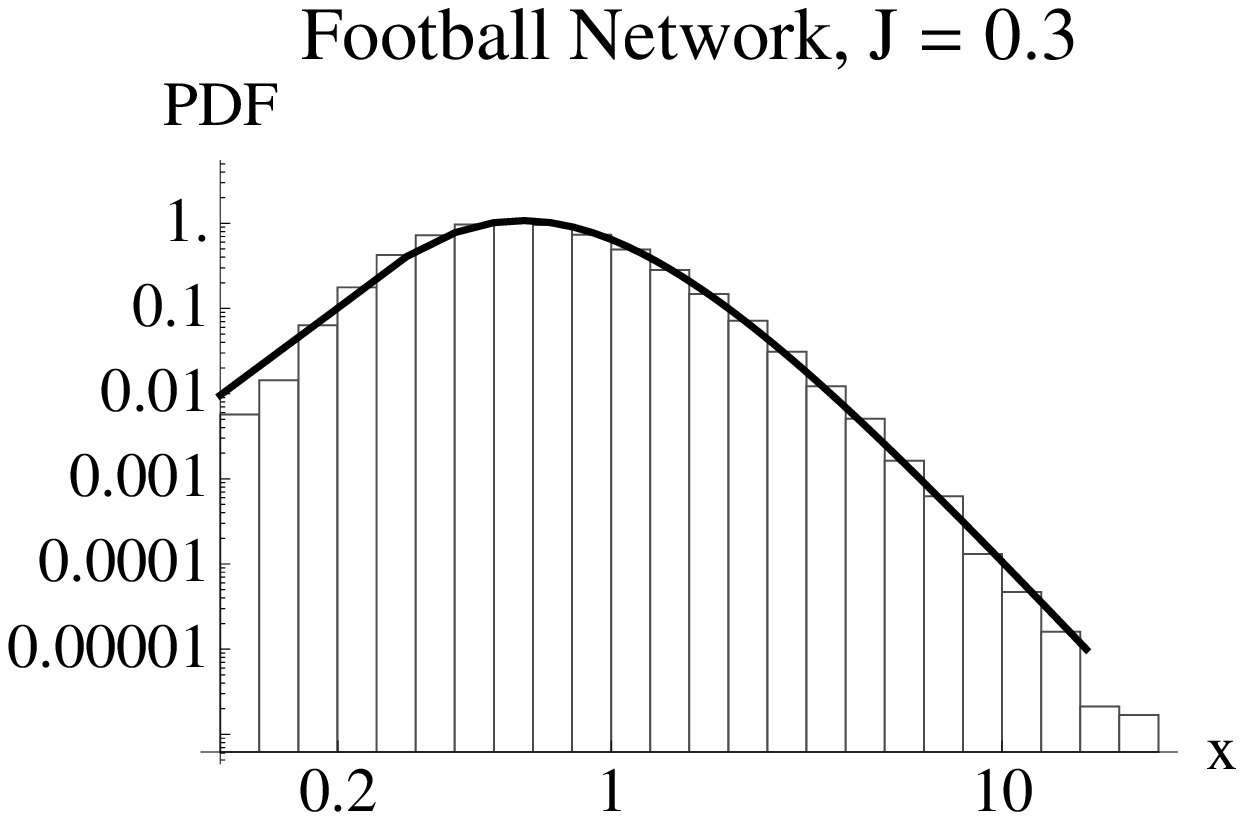}}
 \resizebox{.45\textwidth}{!}{\includegraphics{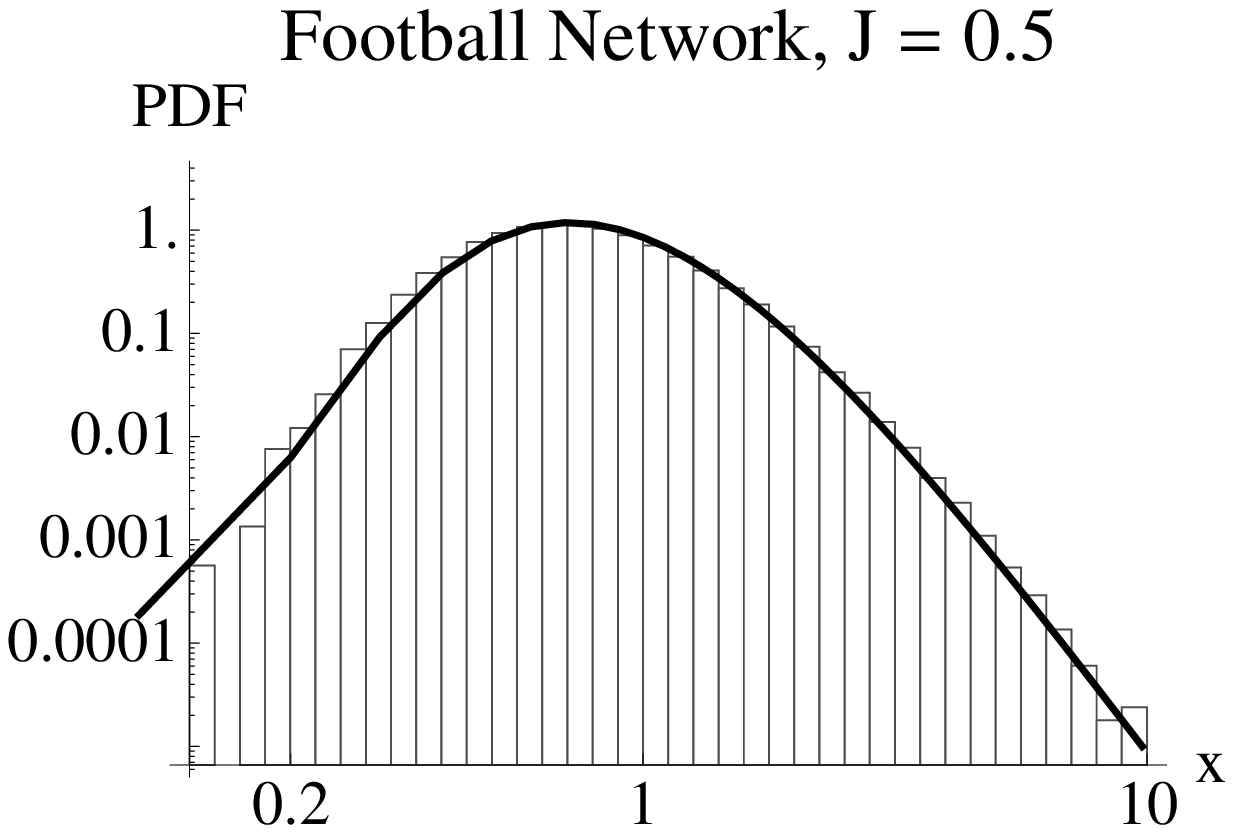}}
 \caption{The simulated PDFs of the BM model on the American college football network for $J=0.3$ and 0.5.
 The solid lines represent the results obtained by our theory.}
 \label{093508_20Aug12}
\end{figure}

The American college football network obtained by Girvan and
Newman\cite{Girvan} includes 115 nodes and 613 edges. The simulated
PDFs for $J=0.3$ and 0.5 are shown in Fig. \ref{093508_20Aug12}. In
both cases, our theory reproduces the PDF well. 

The agreement between theory and simulation seems excellent. We have not
understand the reason for this good agreement, however, one possible
reason is that this network has community structure.
As Girvan and Newman showed,  this  network has clear communities.
As we will see in the later section, the spatial correlation of wealth
would be small for globally coupled network. 
Therefore it is natural to assume that wealth correlation is also small, if
the network is consisted from several densely connected subgraphs.
However, we still need further study to explain this excellent agreement.

\section{Discussion and Conclusions}

Here, we have developed a theory for the BM model on complex
networks. By generalizing our previous work, we were able to propose
a theory applicable to a complex network with a given adjacency
matrix. Using the adiabatic and independent assumptions, we derived
the equations that determine the static PDF of the wealth. The
result was compared to numerical simulations, and we found our
theory works well on a Erd\"os-R\'enyi network, WS networks with $p \ge 0.1$,
and the American college football network. We also found that the
theory did not perform well for a WS network with $p= 0.01$ as the
independent assumption is no longer valid.

The theory does not apply to a WS network with a small $p$ value
because of the large spatial correlation in this case. Thus, we may
ask when and why the spatial correlation becomes large. While it is
difficult to give a complete answer to this question, the following
discussion suggests that the properties of the Laplacian are
essential to this problem.

Suppose that $J$ is very large and the fluctuations of $x$ are very
small. Under such circumstances, we can approximate $xdW$ in Eq.
(\ref{083834_8Aug12}) as $dW$ and obtain
\begin{equation}
 dx_i = - J \sum_j L_{ij} x_j  dt +\sqrt 2 \sigma dW\;,\label{150804_20Aug12}
\end{equation}
where $L=(L_{ij})$ is the Laplacian matrix $L_{ij}= \delta_{ij} d_i
- a_{ij}$. To investigate the correlation in this toy model, we
assume for simplicity that the network is undirected and connected.
Introducing the eigenvalues and corresponding normalized
eigenvectors of $L$ as $\lambda_1 = 0 <  \lambda_2 \le \cdots
\lambda_N$ and $\mathbf{v}_1,\mathbf{v}_2,\cdots \mathbf{v}_N$, we
can write $(x_1, x_2, \cdots , x_N)^T = \sum_{i=1}^N q_i
\mathbf{v}_i$. Then $L$ is diagonalized and Eq.
(\ref{150804_20Aug12}) can be written as
\begin{equation}
 dq_i = - J \lambda_i q_i dt +\sqrt 2 \sigma dW\;.
\end{equation}
The static distribution of $q_i (2\le i \le N)$ is
calculated as
\begin{equation}
 P(q_i) = \frac{\sqrt{J \lambda_i}}{\sqrt{2\pi}
  \sigma}\exp\left[-\frac{J \lambda_iq_i^2}{2\sigma^2}\right]\;.\label{133706_28Aug12}
\end{equation}
The $q_1$ term does not have a static distribution because
$\lambda_1 =0$. However, because $\mathbf{v}_1 =
\frac{1}{\sqrt{N}}(1,1,\cdots,1)$, this term gives the uniform
change, $x_i\rightarrow x_i+C$, and does not contribute to the
correlation.

From Eq. (\ref{133706_28Aug12}) and the relation $x_i = \sum_j q_j
(\mathbf{v}_j)_i$, we obtain the covariance
\begin{equation}
 \langle (x_i-\langle x \rangle)( x_j-\langle x \rangle) \rangle =
  \sum_{l=2}^N \frac{\sigma^2}{J\lambda_l}(\mathbf{v}_l)_i(\mathbf{v}_l)_j\;.\label{145452_24Aug12}
\end{equation}
Our independent assumption implies that the off-diagonal elements of
Eq. (\ref{145452_24Aug12}) are smaller than the diagonal elements.
This condition is satisfied if every $\mathbf{v}_i$  is "localized", in other
words, has only one large components.
It has been shown by some researchers that eigenvectors is localized in many complex networks \cite{McGraw2008,Nakao2010}.
Our theory will be applicable to these network models. 

%For example, we consider the globally coupled network for which
 %$\lambda_2=\lambda_3=\cdots \lambda_N=N$ and $v_l=\frac{1}{\sqrt{N(N-1)}}
 %(1,1,\cdots,-N+1,1,\cdots, 1,1)$. Then, $\sum_{l=2}^N
 %\frac{1}{J\lambda_l} (v_l)_i (v_l)_j = -\frac{1}{JN(N-1)} $ for
 %$i\ne j$, and therefore the covariance is much smaller than the variance $\langle
 %(x_i-\langle x_i \rangle)^2\rangle=\frac{1}{JN}$ for large $N$. This suggests
%that it is the Laplacian rather than the adjacency matrix that is
%important when we consider the dynamics subject to noise.

We also consider here the case for which the adiabatic assumption is
not valid. If the node degree is large enough, then $\tilde x$
changes at a much slower rate than that of $x$. Therefore, this
assumption will be valid if the degree of each node is large. On the
other hand, if the degree of each node is small, then not only the
adiabatic assumption but also the use of the central-limit theorem
cannot be justified. For example, our theory will not work well for
the ``star''-network in which almost all nodes have a degree of 1.

We note that theory for the BM model in the wealth-condensed phase
is still elusive, as the standard central-limit theory is not
applicable when $\langle x^2 \rangle $ diverges. Thus, to discuss
the static properties, we will need to generalize the central-limit
theorem so that it can be adopted even if the variance diverges.

Finally, we note that our developed technique is applicable to other
dynamical systems subject to noise. Stochastic dynamics on complex
network is attracting the interests of many reserchers recently, however,
dynamics with multiplicative noise has not studied so far. 
We believe this method will shed new light on other problems regarding
the dynamics on complex networks.

\end{document}